\documentclass[12pt]{iopart}
\usepackage{cite}

\expandafter\let\csname equation*\endcsname\relax
\expandafter\let\csname endequation*\endcsname\relax
\usepackage{amsmath}
\newcommand{\abs}[1]{\left|#1\right|}
\newcommand{\bra}[1]{\langle #1|}
\newcommand{\ket}[1]{|#1\rangle}
\usepackage{hyperref}

\usepackage{graphicx}
\usepackage{diagbox}
\usepackage{subfig}

\begin{document}


\title[]{Enhanced feature encoding and classification on distributed quantum hardware}

\author{R. Moretti$^{1,2,3}$, A. Giachero$^{1,2,3}$, V. Radescu$^{4}$,  M. Grossi$^{5}$}

\address{$^1$ Department of Physics, University of Milano-Bicocca, Piazza della Scienza, 3, 20126, Milan, Italy}
\address{$^2$ INFN - Milano Bicocca, Piazza della Scienza, 3, 20126, Milan, Italy}
\address{$^3$ Bicocca Quantum Technologies (BiQuTe) Centre, 3, 20126, Milan, Italy}
\address{$^4$ IBM Quantum, IBM Deutschland Research \& Development GmbH - Schöenaicher Str. 220, 71032 Böeblingen, Germany}
\address{$^5$ European Organization for Nuclear Research (CERN), Geneva, CH-1211, Switzerland}

\ead{roberto.moretti@mib.infn.it}
\vspace{10pt}
\begin{indented}
\item[]Month year
\end{indented}

\begin{abstract}
The steady progress of quantum hardware is motivating the search for novel quantum algorithm optimization strategies for near-term, real-world applications. In this study, we propose a novel feature map optimization strategy for Quantum Support Vector Machines (QSVMs), designed to enhance binary classification while taking into account backend-specific parameters, including qubit connectivity, native gate sets, and circuit depth, which are critical factors in noisy intermediate scale quantum (NISQ) devices. The dataset we utilised belongs to the neutrino physics domain, with applications in the search for neutrinoless double beta decay. A key contribution of this work is the parallelization of the classification task to commercially available superconducting quantum hardware to speed up the genetic search processes. The study was carried out by partitioning each quantum processing unit (QPU) into several sub-units with the same topology to implement individual QSVM instances.  We conducted parallelization experiments with three IBM backends with more than $100$ qubits, ranking the sub-units based on their susceptibility to noise. Data-driven simulations show how, under certain restrictions, parallelized genetic optimization can occur with the tested devices when retaining the top $20\%$ ranked sub-units in the QPU.

\end{abstract}

%
\vspace{2pc}
\noindent{\it Keywords}: Quantum Computing, Quantum Feature Maps, Neutrino Physics, Genetic Algorithms, QSVM
%
%
%
%

\section{Introduction}
\label{sec:intro}
The ongoing advancements of quantum computing hardware, in the era of quantum utility \cite{Kim2023, Herrmann_2023}, are driving significant interest in leveraging intermediate-scale quantum devices available today across many research fields. Many research projects focused on finding the potential of Variational Quantum Algorithms (VQAs) in a wide range of fields, including quantum chemistry \cite{vqerev}, combinatorial optimization \cite{comboptrev}, and machine learning. In particular, Quantum Machine Learning (QML) held significant promise by integrating quantum computing into classical machine learning techniques with the hope of achieving better performance in tasks like classification \cite{qmlclassrev}, clustering \cite{Belis2024}, and generative modeling \cite{qmlgenrev}. Yet, despite the high expectations surrounding QML and VQAs, solid theoretical guarantees on their performance and clear demonstrations of quantum advantage remain challenging.


Recent studies underline the need to reconsider the power of QML as a probe for quantum advantage  \cite{9908575, PRXQuantum.3.030101, NEURIPS2021_69adc1e1} or overcome the limitations of early-stage quantum backends \cite{bowles2024betterclassicalsubtleart, Wang_2021, 9908575, Alam2022, Chen2023}, motivating the effort to provide QML algorithm design strategies tailored to the specific hardware in use.

The intrinsic ability of quantum systems to represent data within large Hilbert spaces can lead to highly expressive machine learning models \cite{Peters2021, PhysRevLett.122.040504, PhysRevResearch.3.L032049} that are challenging to simulate using classical methods. For instance, quantum kernel methods, such as Quantum Support Vector Machines (QSVMs) \cite{Havl_ek_2019}, exploit the embedding of input data in a feature space whose dimensionality grows exponentially with the number of qubits in use. Although such dimensionality enhancement can expose generalizability issues \cite{Jerbi2023, Peters_2023, agliardi2024mitigatingexponentialconcentrationcovariant}, this is not necessarily true for any quantum kernel for a given dataset. Also, techniques such as projected quantum kernels \cite{Huang2021, Thanasilp2024} can further help to balance expressivity and generalizability thanks to partial tracing. A remarkable advantage of quantum kernels is that, like their classical counterpart, they are positive and semi-definite functions of the input features, simplifying the training process of (Q)SVMs by making it a convex optimization problem \cite{Cortes1995}.

The standard quantum kernel formulation is referred to as ``fidelity quantum kernel'' \cite{Huang2021, schuld2021supervisedquantummachinelearning}, defined as
\begin{equation}
\label{eq1}
    \mathcal{K}: \mathcal{K}(\vec{x}_i, \vec{x}_j) \rightarrow \abs{\bra{\phi(\vec{x}_i)}\ket{\phi(\vec{x}_j)}}^{2},
\end{equation}
where $\vec{x}_i$ and $\vec{x}_j$ are feature vectors with $i, j$ corresponding to sample indices in the dataset and $\phi$ is an arbitrary feature map that encodes input features in a (multi-)qubit state. We can rewrite Eq.~\ref{eq1} more practically in terms of a unitary ansatz that implements the transformation $U\ket{0}^{\otimes{n}} = \ket{\phi}$:
\begin{equation}
\label{eq2}
    \mathcal{K}: \mathcal{K}(\vec{x}_i, \vec{x}_j) \rightarrow \abs{\bra{0}^{\otimes{n}}U(\vec{x}_i)^{\dag} U(\vec{x}_j) \ket{0}^{\otimes{n}}}^{2}.
\end{equation}
The performance of the resulting QSVM depends significantly on the chosen ansatz and identifying an optimal quantum feature map for a given dataset remains an open challenge in the field. This task is often approached through manual exploration of various heuristic feature maps \cite{Vasques2023, Li2023}, kernel alignment techniques \cite{Cristianini2006}, or incorporating trainable parameters to the unitary operator $U$ \cite{Hubregtsen_2022, PhysRevApplied.21.054056, rodriguezgrasa2024trainingembeddingquantumkernels}. These approaches are either time-consuming or expose to trainability hurdles such as the occurrence of barren plateaus \cite{McClean2018, Cerezo2021} which are typical of Parametrized Quantum Circuit (PQC).

An important related challenge is the phenomenon of exponential concentration \cite{Thanasilp2024, Slattery_2023}, where highly expressive kernels exhibit an exponential concentration of kernel values due to the high dimensionality of the mapped feature vectors, distributed \emph{uniformly} across the exponentially large Hilbert space. As a result, distinguishing between kernel values requires an exponentially large number of measurements. This issue becomes more severe when considering Noisy Intermediate-Scale Quantum (NISQ) devices \cite{Preskill_2018, RevModPhys.94.015004}, where the precision of quantum kernel evaluations is significantly degraded by noise. While it is demonstrated that the projected quantum kernel techniques mentioned earlier help reduce exponential concentration, specific fidelity quantum kernel optimization strategies may also effectively address this problem.

Recent efforts have focused on automating the optimization of quantum feature maps at the gate level, with genetic algorithms emerging as promising tools for this purpose \cite{AltaresLopez2021, ALTARESLOPEZ2024122984, PellowJarman2024, PhysRevResearch.5.013211, Wang2024, wang2023fitnessfunctionsentanglementgates, creevey2023kernelalignmentquantumsupport, incudini2023automaticeffectivediscoveryquantum}. Genetic ansatz optimizations have also been carried out in the context of Quantum Approximate Optimization Algorithm (QAOA) \cite{schiavello2024evolvingmultipopulationevolutionaryqaoadistributed}. However, current implementations have primarily demonstrated proof-of-concept results and have not fully addressed the reproducibility of quantum kernels on NISQ devices. Moreover, most of these optimization strategies \cite{AltaresLopez2021, ALTARESLOPEZ2024122984, PellowJarman2024, PhysRevResearch.5.013211, Wang2024, wang2023fitnessfunctionsentanglementgates, creevey2023kernelalignmentquantumsupport} have been tested on standard, well-known datasets with minimal class overlap, achieving near-perfect accuracy with small training sets -- an unrealistic scenario for many practical applications.

\noindent Our study aims to fill these gaps by designing a genetic algorithm that optimizes quantum feature maps while taking into consideration specific hardware characteristics, with their native gates and qubit connectivity map in an extended setting using multiple chips, targeting quantum utility implementation.
We consider a dataset taken from the neutrino physics domain consisting of a Monte Carlo simulation of three-dimensional tracks produced by ionization of liquid argon in a Liquid Argon Time Projection Chamber (LArTPC) detector \cite{Moretti2024, Bonivento:2024qpn}. The physics motivation is related to the possibility of assessing the sensitivity of next-generation experiments like DUNE to the discovery of one of the most elusive processes in low-energy physics, i.e.\ the neutrinoless double beta decay of ${}^{136}$Xe \cite{PhysRevD.106.092002, dunecollaboration2024dunephaseiiscientific}. Crucially, we expect this dataset to have considerable class overlap, hence being a hard binary classification task. The structure of the paper follows the pipeline of the algorithm: after applying an autoencoder-assisted feature preprocessing, which is detailed in Sec.~\ref{sec:preprocessing}, we introduce a genetic optimization strategy that enhances the kernel  classification accuracy on noisy hardware in Sec.~\ref{sec:genetic}. In Sec.~\ref{sec:statevec} we describe the results obtained by the genetic optimizations with a statevector simulation. The simulation under ideal conditions represents a necessary step to effectively run the search for optimal quantum kernels on the quantum computer.

\begin{figure}
    \centering
    \includegraphics[width=0.9\linewidth]{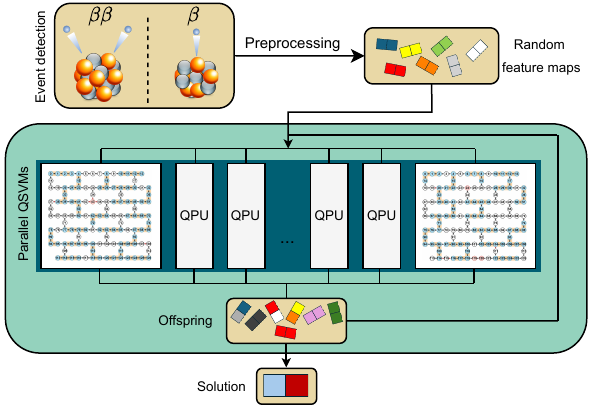}
    \caption{Schematic workflow proposal for the automatic quantum feature map optimization via genetic algorithm for classifying $\beta$ and $\beta\beta$ topologies. Parallelization is achieved through QPU partitioning and leveraging each partition to compute several quantum kernels simultaneously. Additional speedup can be provided by splitting the task through multiple QPUs.}
    \label{fig:workflow}
\end{figure}

\noindent Finally, to make the most of quantum computational resources for an utility scale experiment, we assess the feasibility of parallelizing IBM quantum chips for computing multiple quantum kernels simultaneously, by partitioning three quantum processing units (QPU) into several, independent sub-units. We evaluate the impact of noise and the robustness of this parallelization approach in Sec.~\ref{sec:experiment}, where we show the potential for accelerating genetic optimization by computing entire generations simultaneously, with a single quantum circuit containing all quantum kernels within a generation. Fig.~\ref{fig:workflow} depicts the workflow proposed in this study. This technique could demonstrate a practical, real-world utility-scale application for currently available QPUs, driving progress in data analysis across fields of study. We summarize our findings and potential future directions in Sec.~\ref{sec:conclusions}.

\section{Physical motivation and autoencoder processing}
\label{sec:preprocessing}
One of the most important physics processes under investigation within the neutrino physics community is the neutrinoless double beta decay ($0\nu\beta\beta$) \cite{0nubb, Agostini_2023}. This process can be conceptually understood as a nucleus undergoing double beta decay, where the two (anti-)neutrinos normally emitted during beta decays annihilate each other, leaving only two electrons in the final state and the modified nucleus. Demonstrating the existence of such a process would have considerable implications for our understanding of Fundamental Physics, the matter-antimatter asymmetry in the Universe \cite{PhysRevD.102.055009} and determine whether neutrinos and antineutrinos correspond to the same particle \cite{PhysRevD.108.095028}. A number of studies suggest that ton-scale Liquid Argon Time Projection Chambers under certain circumstances can exhibit high discovery potential for the $0\nu\beta\beta$ decay of the ${}^{136}$Xe \cite{PhysRevD.106.092002, Andringa2023, Bezerra2023}, comparably to other experiments that have been specifically designed for this task, and currently pursued \cite{next, kamland-zen, panda, Agostini2020, Adhikari2022}. Such LArTPCs are currently being developed within the Deep Underground Neutrino Experiment (DUNE), and the search for the $0\nu\beta\beta$ decay (together with the study of other few-MeV scale events) will be potentially taken into consideration in the realization of Dune Phase II Far Detector LArTPCs, called Module of Opportunity (MoO) \cite{dunecollaboration2024dunephaseiiscientific}.

\noindent The ${}^{136}$Xe $0\nu\beta\beta$ search with a LArTPC leverages the fact that xenon-doping of liquid argon will be pursued, because it will boost the scintillation light detection efficiency emitted by argon \cite{SotoOton2022, verticaldrift, dopingxe}, improving the overall detection performance. Assuming xenon-doping occurs with xenon enriched in the ${}^{136}$Xe isotope, we expect the decay to happen inside the active volume, with the signature consisting of two electrons ranging from the same point in space, producing a single ionization track. TPCs allow us to measure the total energy produced in the decay, and also reconstruct the full three-dimensional track information as a register of ``hits''. Based on that, to boost the experimental sensitivity, we can develop a classification strategy leveraging the different topologies of signal events ($\beta\beta$ decays) and one of the most problematic background channels for this analysis, which is the $\beta$ decay of ${}^{42}$Ar \cite{PhysRevD.106.092002, Moretti2024}.\\
Considering a realistic granularity for a ton-scale LArTPC and the average track length produced by this decay in liquid argon ($\sim 1.2$ cm) \cite{Berger2017}, most of the track information will be lost, with a detrimental effect on the classification performance. This classification task is therefore expected to be hard, making it an intriguing test bench for QSVMs, not only because it relates to a real-world application, but also because we anticipate highly overlapping (and possibly nontrivially correlated) feature distributions.

\noindent The dataset consists of an ensemble of labeled $0\nu\beta\beta$ events ($\beta\beta$ topology) and single $\beta$ decay ($\beta$ topology), assuming a resolution of $[2\times2\times1]\,\text{mm}^3$ and a hit energy threshold of $50$ keV. Deep learning approaches consisting of a Convolutional Neural Network and a Transformer, when trained with more than $10^5$ samples have reached around 76\% accuracy for the same dataset and the same experimental settings considered in this study \cite{Moretti2024}.

\subsection{Autoencoder feature extraction}
\begin{figure}
    \centering
    \includegraphics[width=\linewidth]{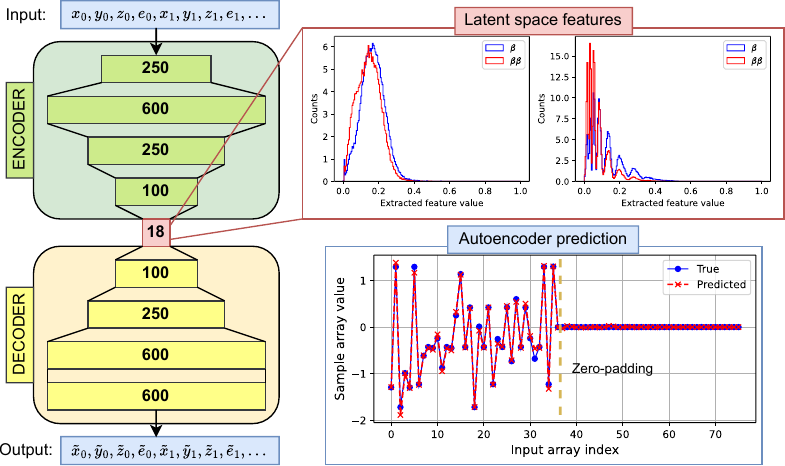}
    \caption{Autoencoder architecture used to extract a latent space representation of the input track information. The input array, as a set of concatenated hit information, is fed through several fully connected layers whose neuron number is indicated in the corresponding block in the diagram. In the top-right frame, two of the $18$ latent space feature distributions are shown within classes. A comparison between the input and the autoencoder output for a single track is reported in the bottom-right frame.}
    \label{fig:autoencoder}
\end{figure}

Each event in the dataset consists of an ensemble of hits produced inside the detector, with each hit carrying four scalar quantities (three spatial coordinates and hit energy). The number of hits varies for each different track, making the pre-processing not trivial. The adopted solution is the autoencoder, whose architecture can handle this flexibility while performing the data reduction as a label-agnostic step.
The autoencoder consists of a multilayer perception \cite{MURTAGH1991183} that learns an identity function of the input through one or more intermediate layers whose dimension is smaller than the input size. The number of reduced features in the ``latent space'' is controlled by the size of the smallest layer, from which these features are extracted. In this study, we determined the optimal compression of the original representation, reducing its size from 
$76$ to $18$, based on the autoencoder's loss function and latent space dimensions. This compression significantly simplifies statevector QSVM simulations while minimizing information loss. The autoencoder architecture, feature distributions, and a sample prediction are illustrated in Fig.~\ref{fig:autoencoder}.

We designed and trained the autoencoder with the TensorFlow package \cite{tensorflow2015-whitepaper} by partitioning the dataset into $120\times10^3$ training and $30\times 10^3$ validation samples, minimizing the Mean Square Error cost function between the input data and the network prediction using the Adam stochastic gradient descent algorithm \cite{KingBa15}. Input data is passed to the autoencoder as a single array for each sample, by concatenating hit arrays $(\vec{h}_0,\, \vec{h}_1,\, ...\,,\, \vec{h}_n)$ where $n$ is the hit number for a particular event, and $\vec{h}_i = (x_i,\, y_i\, z_i, \, e_i)$, i.e.\ three spatial and one energy coordinates. The input is zero-padded so that every track is associated with an array of length $76$. We observe how two example latent space feature distributions shown in Fig.~\ref{fig:autoencoder} are irregular, either skewed or multi-peaked, with significant class overlap as we expected from the dataset. The features were rescaled in the interval $[0, 1]$ and used as input to the quantum kernels described in Sec.~\ref{sec:statevec} and Sec.~\ref{sec:experiment}.

\section{Genetic algorithm implementation}
\label{sec:genetic}
The genetic optimization implemented in this study aims to lead to the generation of quantum kernels that,  at the same time, allow high classification accuracy with resource-efficient circuit embeddings \cite{wang2024powercharacterizationnoisyquantum}. To this extent, we employ a gradient-free method that optimises complex (generally non-differentiable) structures such as gate types, rotation angles, and qubit connectivities. Genetic algorithms are powerful tools that serve our purpose. The Genetic algorithm is a meta-heuristic method for solving optimization problems inspired by biological evolution processes \cite{Katoch2021}. In particular, given a fitness function that quantifies the goodness of a solution through one or more objectives, the algorithm aims to generate solutions that maximize the fitness function output (if mono-objective) or optimal solutions in terms of multi-objectives trade-off \cite{incudini2023automaticeffectivediscoveryquantum}. The solutions are generally referred to as ``chromosomes'', and in our study, each chromosome corresponds to a quantum feature map for the kernel. Each chromosome consists of an ensemble of ``genes'' that encode the chromosome properties. The genetic content of chromosomes can be modified through operations called ``mutation'' and ``crossover''. Mutation is a one-to-one operation that takes in input a chromosome and returns an offspring, in which a fraction of genes is replaced with a random value. Crossover is a two-to-one operation that takes as input two chromosomes and combines their genetic content into an offspring solution. Despite these operations being fundamental in providing the necessary exploration power throughout the genetic search, they require careful calibration so as not to destroy advantageous chromosome characteristics with too high probability. Excessive mutation and crossover rate in the genetic algorithm would lead to approximate the behaviour of a mere random search.

\noindent Our specific genetic optimization integrates the PyGAD library \cite{Gad2024} for gene handling and the Qiskit software \cite{qiskit2024} for computing the quantum kernels. Our algorithm follows from a standard iterative procedure and can be summarized as follows:
\begin{enumerate}
    \item Initialization: The algorithm begins by initializing a population of $N$ randomly generated chromosomes. The population size remains constant across generations, setting an upper limit on the number of solutions explored in each generation. Mutation and crossover rates ($\mu$ and $\gamma$, respectively) are also fixed at this stage, as well as the definition of fitness function.
    \item Chromosome evaluation and parent selection: Each chromosome in the population is evaluated based on its fitness function output. A subset of the population, consisting of $M$ chromosomes, is selected to survive into the next generation as parents without gene alteration. We based our selection criteria on a steady-state selection \cite{Agapie2014} when considering a mono-objective fitness function and on the Non-dominated Sorting Genetic Algorithm II (NSGA-II) \cite{nsgaii} when considering a multi-objective one. These methods assign a probability to each chromosome of being retained for the following generation. In some cases, we alternatively recur to an elitism strategy, in which the best $K$ chromosomes in the generation are kept as parents with probability one, leaving no other parent to be selected. 
    \item Offspring generation: The remaining $N-M$ chromosomes in the new generation (or $N-K$ when elitism is applied) are generated as offspring. These are produced by combining the genetic information from the $M$ selected parents through a two-point crossover \cite{Umbarkar2015} and subsequent mutation operations. For each offspring, a random pair of parents is selected, with crossover occurring with a probability $\gamma$. If crossover does not occur, the offspring is a direct copy of the first parent. Each offspring chromosome then undergoes mutation (regardless crossover happened or not), where a fraction $\mu$ of its genes are randomly replaced.
    \item Steps (ii) and (iii) are repeated until a stop criterion is met. This criterion could be a predefined maximum number of generations or a convergence condition on the fitness function values within generations.
\end{enumerate}

\subsection{Gene representation}
To allow the genetic algorithm to explore a wide range of feature maps throughout generations, mutation and crossover operations should be able to affect the quantum feature map circuit in the most general way possible. To achieve this, we translate the quantum circuit structure into a convenient array of integer numbers (genes) that encode specific pieces of information according to their value and position in the array, such as gate type and order. In this gene representation, rotation angles for variational gates are included, as well as specific gene-dependent transformations of one or two dataset features.

We associate a sub-array of six integers for each gate composing the circuit. The position of each sub-array inside the whole gene array determines the order in which it is applied, as well as the specific qubit(s) involved in the gate. The gate-gene correspondence is summarized in Table~\ref{tab:genemap}, and the gene-dependent gate arguments are collected in Table~\ref{tab:genemap2}. The gate type gene defines whether the other genes will have an effect on the quantum circuit or not. For instance, if the gate type is non-parametric, the other genes for that gate referring to features and transformations will be ignored.

\begin{table}
\centering
\begin{tabular}{|p{4cm}|p{2cm}|p{2cm}|p{6cm}|}
\hline
\textbf{Gene} & \textbf{Position} & \textbf{Range} & \textbf{Description} \\ 
\hline
Gate type           & 0 & [0, $N$)  & Identifies a gate type among the allowed ones. \\
\hline
Transformation (T)      & 1 & [0, 2]                            & Determines a transformation type.\\
\hline
Multi-feature (MF)      & 2 & [0, 1]                            & Either using one or two features in the gate argument. \\
\hline
First feature index & 3 & [0, F]                           & Feature index to be used as first rotation argument. \\
\hline
Second feature index & 4 & [0, F]                          & Feature index to be used as second rotation argument \\
\hline
Second qubit index  & 5 & [0, $Q$) - \{q\}             & Target qubit for the two-qubit gate. \\
\hline
\end{tabular}
\caption{Genes expressing a single gate in the feature map circuit acting on a specific qubit $q$ in the circuit. If the gate type gene is associated with a two-qubit gate, $q$ corresponds to the control qubit, while the target qubit is specified by the second qubit index gene. Each gene is an integer number bound in specific intervals, depending on the number of allowed gates N, total features in the dataset F, and qubits in the quantum circuit Q.}
\label{tab:genemap}
\end{table}

\begin{table}
\centering
\renewcommand{\arraystretch}{1.3} 
\setlength{\tabcolsep}{6pt}
\begin{tabular}{|c|>{\centering\arraybackslash}m{6cm}|>{\centering\arraybackslash}m{6cm}|}
\hline
\backslashbox{T}{MF} & \textbf{0} & \textbf{1} \\ 
\hline
\textbf{0} & $2\pi (x_i - 0.5)$ & $2\pi x_i(1-x_j) - \pi$ \\ 
\hline
\textbf{1} & $2\pi (x_i - 0.5)(1-x_i) - \pi$ & $(2 \pi (x_i - 0.5)(1-x_j) - \pi)\;\times$ \newline $(2\pi (x_j - 0.5)(1-x_i) - \pi)) /\pi$ \\ 
\hline
\textbf{2} & $2\arcsin(2x_i - 1) - \pi$ & $2\arcsin((2x_i - 1)(2x_j - 1)) - \pi$ \\ 
\hline
\end{tabular}

\caption{Mathematical expressions for gate arguments determined by the Transformation (T) and Multi-feature (MF) genes. Each expression is bound between either $[-\pi, \, \pi]$ or $[0, -2\pi]$ for $x_i,\, x_j \in [0, 1]$. $i$ and $j$ correspond to the value of the first and second feature index genes respectively, as referred to in Table~\ref{tab:genemap}.
}
\label{tab:genemap2}
\end{table}

To roughly set a circuit size for the quantum kernels, we define an initial circuit size parameter $S$, such that the gene array length is $6\times Q\times S$. This represents the total number of single-qubit gates acting on one qubit, plus the number of two-qubit gates where that qubit serves as the control qubit. We note that the gate number of a feature map can be smaller than $Q\times S$ when the identity gate is allowed in the circuit, making it possible for quantum feature maps to evolve into simpler circuits when needed.

\subsection{Fitness function choices}
\label{ssec:fitfun}
By considering quantum kernels evaluated through statevector simulations, we identify three QSVM metrics to compose one mono-objective and one multi-objective fitness function: 
\begin{itemize}
    \item $a$ - k-fold cross-validation accuracy of the QSVM's predictions, where the accuracy is evaluated as the fraction of correctly predicted samples over the total.
    \item $\sigma$ - standard deviation of the off-diagonal kernel matrix entries.
    \item $d$ - feature map circuit depth, which can be defined by the maximum path length (number of gates) connecting one qubit in the initial state and one in the final state, moving in the forward direction only. An equivalent definition in terms of complexity time steps is provided in \cite{Gyongyosi2020}.
\end{itemize}
While promoting QSVMs with the highest $a$ is of primary importance, this metric alone is insufficient in NISQ setting where the quantum kernel matrix cannot be evaluated with arbitrary precision. The second and third metrics under consideration for this study allow us to estimate the feasibility of implementing a feature map on quantum hardware, without depending on a specific noise model.
$\sigma$ is an intuitive -- yet powerful-- metric that is correlated to the precision required by quantum hardware in estimating the kernel matrix. The ability of quantum hardware to distinguish between two different entries of the kernel matrix depends on both the number of measurement shots and the level of noise. An increased $\sigma$ corresponds to a lower precision requirement, as the difference between kernel entries will generally be higher.
In addition, promoting feature maps leading to high $\sigma$ will also mitigate the exponential concentration problem, which manifests in a rapid decrease in $\sigma$ as the number of qubits and circuit depth grow, without relying on projected quantum kernels. The circuit depth $d$ straightforwardly addresses the quantum kernel estimation feasibility on noisy quantum hardware, as a higher depth implies more gates to be implemented sequentially, increasing the computation time and, therefore, the coherence time requirements for the qubits in the device. The circuit depth metric, being dependent on particular native gate sets and qubit connectivities, allows us to promote quantum circuits that are tailored to a specific QPU design.


\noindent We consider relying on statevector simulation and the control metrics mentioned above as the most efficient way to rapidly evaluate as many quantum kernels as required for the genetic search. Other approaches may involve leveraging noise models in the quantum kernel estimation, or real QPUs directly. This would partially replace the need for $\sigma$ and $d$. However, these methods would be either computationally expensive or require excessive runtime for modern QPU availabilities, at this preliminary stage of the investigation. QPU parallelization, as motivated in Sec.~\ref{sec:intro} and investigated in Sec.~\ref{sec:experiment}, would overcome the second obstacle.

\section{Statevector genetic optimization}
\label{sec:statevec}
We carried out several genetic optimizations for several initial settings, classically simulating from $4$ to $12$ qubits quantum kernels on a statevector backend. We randomly extracted a subset of $500$ samples from the original dataset and employed it for all the analyses described in this section. Every run was initialized with $N=50$ feature maps by generating chromosomes whose gene content was uniformly sampled from the intervals reported in Table~\ref{tab:genemap}. We let the feature maps evolve until completing $250$ generations, which always resulted in reaching a fitness function convergence, and resorted to elitism by setting $K=5$, meaning that the best $5$ QSVMs in each generation (according to either the steady state selection or the NSGA-II algorithm) were propagated to the following one without any change in their genes. Running a genetic algorithm with these settings requires computing a total of $N+250(N-5)=11300$ distinct quantum kernels.

\noindent Considering the fitness function metrics discussed in Sec.~\ref{ssec:fitfun}, we opted to design two distinct fitness function categories. The first one was mono-objective and consisted of a weighted sum of five-fold cross-validated accuracy $a$ and the standard deviation of the kernel matrix off-diagonal elements $\sigma$:
\begin{equation}
    f=a+ \eta \sigma.
    \label{eq:mono}
\end{equation}
The underlying idea was to keep $\eta$ low enough to mainly promote QSVM with high accuracy, while preferring the kernel matrix with the highest variability only when the accuracies were equal or very similar. The value of $\eta = 0.025$ was determined experimentally.
\begin{figure}
    \subfloat[$4$-qubits fitness function through generations via steady-state selection.\label{fig:statevec1a}]{
        \includegraphics[width=0.47\linewidth]{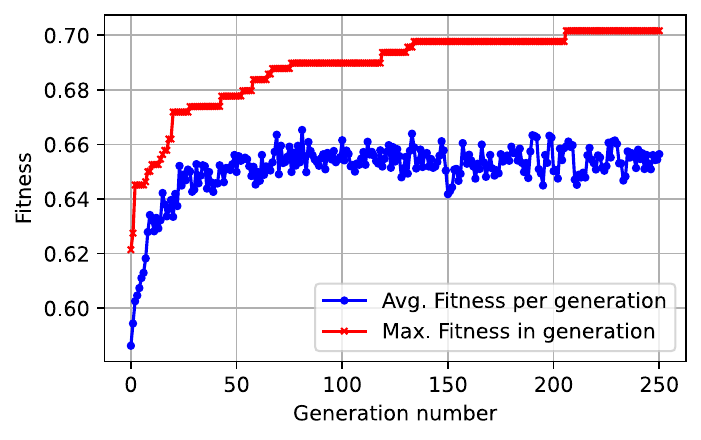}
    }
    \hfill
    \subfloat[Distribution of $4$-qubits QSVM performances in the $\left(a,\,\sigma\right)$ plane.\label{fig:statevec1b}]{
        \includegraphics[width=0.47\linewidth]{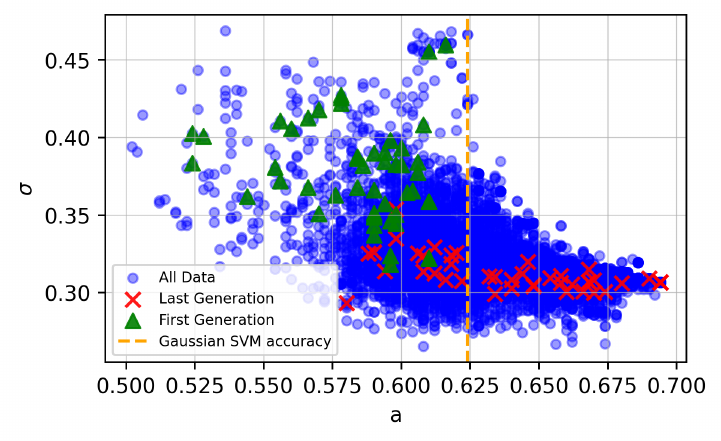}
    }
    \vspace{0.5cm}

    \subfloat[$12$-qubits fitness function through generations via steady-state selection.\label{fig:statevec1c}]{
        \includegraphics[width=0.47\linewidth]{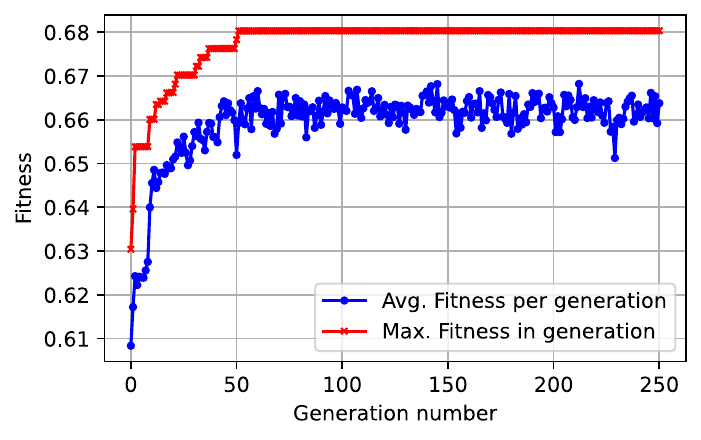}
    }
    \hfill
    \subfloat[Distribution of $12$-qubits QSVM performances in the $\left(a,\,\sigma\right)$ plane.\label{fig:statevec1d}]{
        \includegraphics[width=0.47\linewidth]{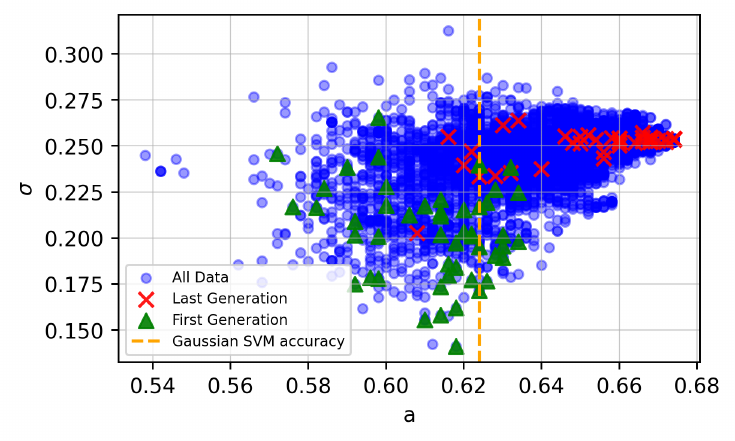}
    }
    \caption{Optimized genetic feature maps for $\beta$ and $\beta\beta$ classification are simulated with a statevector backend. Fitness function values, defined in Eq.~\ref{eq:mono}, show improvement across generations for average (blue line) and best (red line) in (a) and (c). Cross-validation accuracy and kernel matrix standard deviation distributions in (b) and (d) compare the initial (green) and final (red) QVSMs, also benchmarked against a classical out-of-the-box SVM (vertical orange dashed line). Allowed gates: I, H, X, SX, RX, RY, RZ, CX, CRX, CRY, CRZ. Circuit size: $S=8$.}
    \label{fig:statevec1}
\end{figure}
In Fig.~\ref{fig:statevec1} we report the two extreme cases $4$ and $12$ qubits in which we can clearly observe a positive trend in the fitness function throughout the generations, both for the optimal kernel in each generation and the average population fitness. This trend is also represented as the difference between the first generation kernels -- that were randomly selected -- and the last generation in the $\left(a,\,\sigma\right)$ plane. In Fig.~\ref{fig:statevec1b} and~\ref{fig:statevec1d}, we marked the highest cross-validated accuracy achieved with our attempts to train classical SVM, using the scikit-learn package \cite{scikit-learn}, represented by a Gaussian kernel. The Gaussian SVM hyperparameters have been optimized through a grid-search technique, leading to $a=62.5\, \%$. This SVM is outperformed on the same dataset by many QSVMs, peaking at $69.5\, \%$ for $4$ qubits, and $67.5\, \%$ for $12$ qubits. Interestingly, the $4$-qubits search leads to higher fitness values than in the $12$-qubits case when considering the best kernel in each generation, while the contrary can be observed for the population average. We explain these effects by considering two key differences arising from this kind of optimization strategy. First, the number of genes increases linearly with the number of qubits, exponentially increasing the space of possible kernels and slowing down the genetic search. Second, the number of features ($18$) is comparable with the number of variational quantum gates composing the quantum circuits. With the specific parameters used in this study, the average number of variational gates in the feature map is $18$ for the $4$-qubit case and $52$ for the $12$-qubit case. Therefore, the probability of missing an informative feature for classification in a $12$-qubit kernel circuit is lower, leading to higher accuracy on average.
It is important to note that these considerations depend on the initial genetic settings and can be mitigated by increasing the parameter $S$ and varying the allowed gates in the circuits.

\begin{figure}
    \centering
    \subfloat[$4$-qubits fitness function objectives optimization via NSGA-II.]{
        \centering
        \raisebox{15.6mm}{ \includegraphics[width=0.47\textwidth]{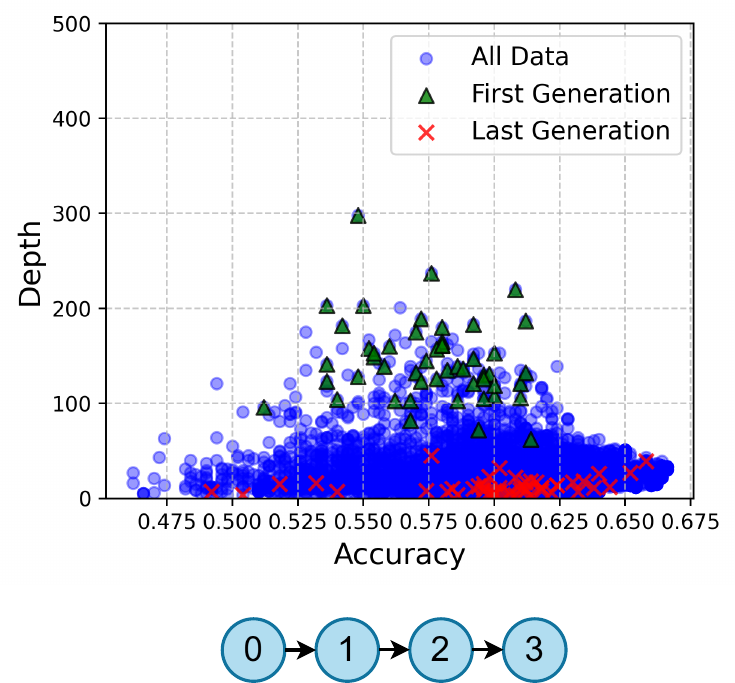}}
        \label{fig:4qb_nsgaii}
    }
    \hfill
    \subfloat[$12$-qubits fitness function objectives optimization via NSGA-II.]{
        \centering
        \includegraphics[width=0.47\textwidth]{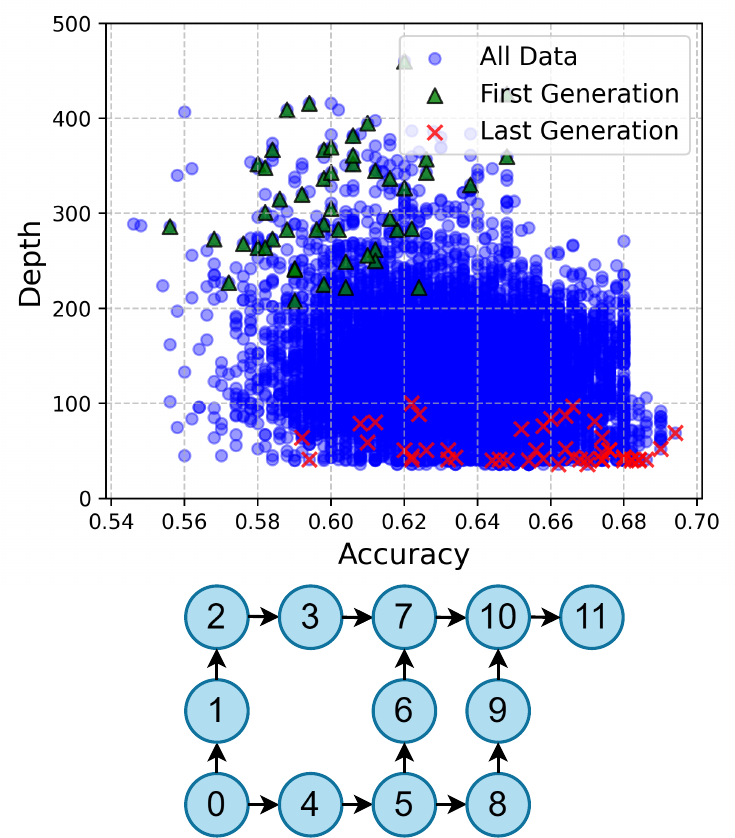}
        \label{fig:12qb_nsgaii}}
    
    \caption{Optimized genetic feature maps for $\beta$ and $\beta\beta$ classification are simulated via a statevector backend without noise. (a) and (b) show the distributions of transpiled circuit depth and cross-validation accuracy, comparing the initial (green) and final (red) QVSMs under specific qubit connectivity assumptions. The fitness function is defined in Eq.~\ref{eq:multi}. Allowed gates: I, H, X, SX, RX, RY, RZ, CX, ECR, CRX, CRY, CRZ. Circuit size: $S=8$. Arrows in the connectivity maps represent the operation direction (control $\rightarrow$ target) for the ECR gate.}
    \label{fig:nsgaii}
\end{figure}

\noindent The second type of fitness function under investigation was multi-objective and takes the form:
\begin{equation}
f=
\begin{pmatrix}
    a \\
    -d \\
\end{pmatrix}.
\label{eq:multi}
\end{equation}
Hence we performed the genetic search through the Non-dominated Sorting Genetic Algorithm II (NSGA-II) to optimize the trade-off between accuracy and circuit depth. The results are shown in Fig.~\ref{fig:nsgaii}. We simulated a $4$-qubit system with a spin-chain connectivity map and a $12$-qubit system with a connectivity map as depicted in Fig.~\ref{fig:12qb_nsgaii}. The circuit depths are calculated on a transpiled circuit considering the respective connectivity maps and a basis gate set consisting of I, X, SX, RZ, ECR.

In this case, we observe how the feature maps evolve towards optimal configurations in terms of high accuracy and low depth. In the $4$-qubits case, we observe how accuracy in the population is generally lower than in the mono-objective optimization case, possibly suggesting that under a certain level of complexity of the quantum circuit, the QSVMs expressivity starts to decrease. The same observation does not apply to the $12$-qubit case, suggesting that the Hilbert space accessed by $12$ qubits is enough, on average, to prevent expressivity loss.

\section{Parallelization of intermediate-scale quantum hardware}
\label{sec:experiment}
Despite the superconducting platform being significantly faster than other quantum computing platforms, such as trapped ions or neutral atoms\cite{superconductingvstrapped, Wintersperger_2023}, computing a full genetic optimization run as in Sec.~\ref{sec:statevec} using cloud-based access to commercially available superconducting quantum backends would require several days of continuous run time. Computing one single kernel matrix element requires repeated executions (on the order of thousands) of the quantum kernel circuit, for just one element in the population. To mitigate this technological bottleneck we propose a partition of the quantum processing unit to parallelize the algorithm execution taking advantage of the large number of available qubits, executing the calculation for all kernels within the same calibration window (intra-chip execution). Moreover, we propose a multi-device parallelization, exploiting modular execution for chips of the same hardware release (intra-platform execution). This allows us to simultaneously extract the measurement outcome from each independent sub-unit. In this section, we describe in detail the parallelization routine for $4$-qubits quantum kernels evaluation, assessing the impact of noise for different backends.

\noindent To assess the performance of intra vs extra platform execution we ran the experiment on three different superconducting backends provided by IBM: ``ibm\_nazca'' (127 qubits), ``ibm\_brisbane'' (127 qubits) and ``ibm\_torino'' (133 qubits), considering a subset of $100$ samples extracted from the original dataset. In the comparisons across the different backends (extra-platform), it was of fundamental importance to consider the different native gate sets for each QPU in choosing the circuit to run and identifying the qubit partitions. ibm\_nazca and ibm\_brisbane share the same basis gates (I, RZ, SX, X, ECR), using the echoed cross-resonance (ECR) gate as a native two-qubit interaction. For these devices, the ECR gate acts in one direction, i.e.\ for each qubit pair in the QPU, only one qubit can be designated as control, and only the other one as target for the gate. Despite sharing similar connectivity maps, ibm\_nazca and ibm\_brisbane have different target-control ordering, requiring different partitioning to compare their performances.

\begin{figure}
    \centering
    \subfloat[ibm\_nazca.]{
        \centering
        \includegraphics[width=0.47\linewidth]{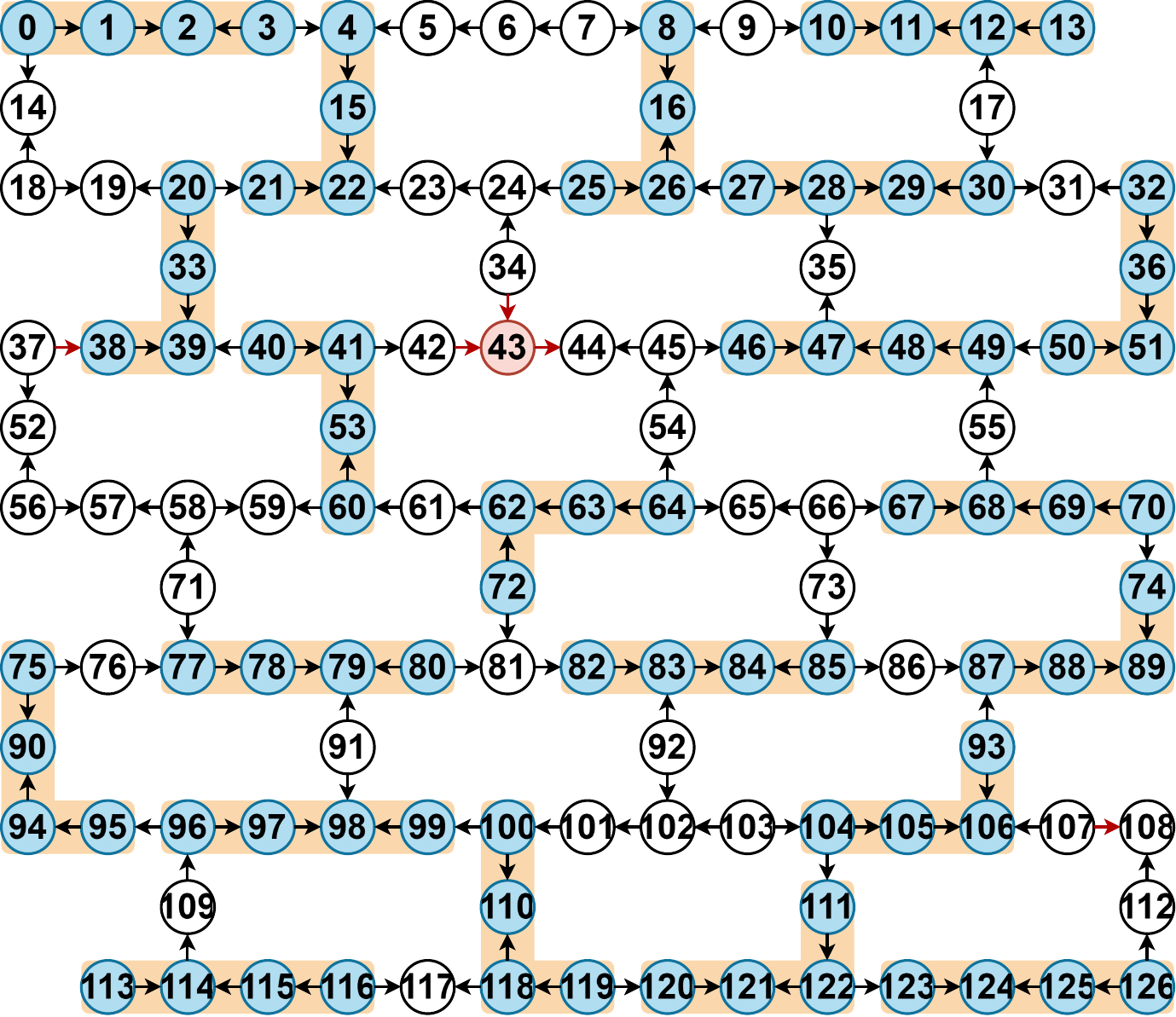}
        \label{fig:nazca_partition} }
    \hfill
    \subfloat[ibm\_brisbane.]{
        \centering
        \includegraphics[width=0.47\linewidth]{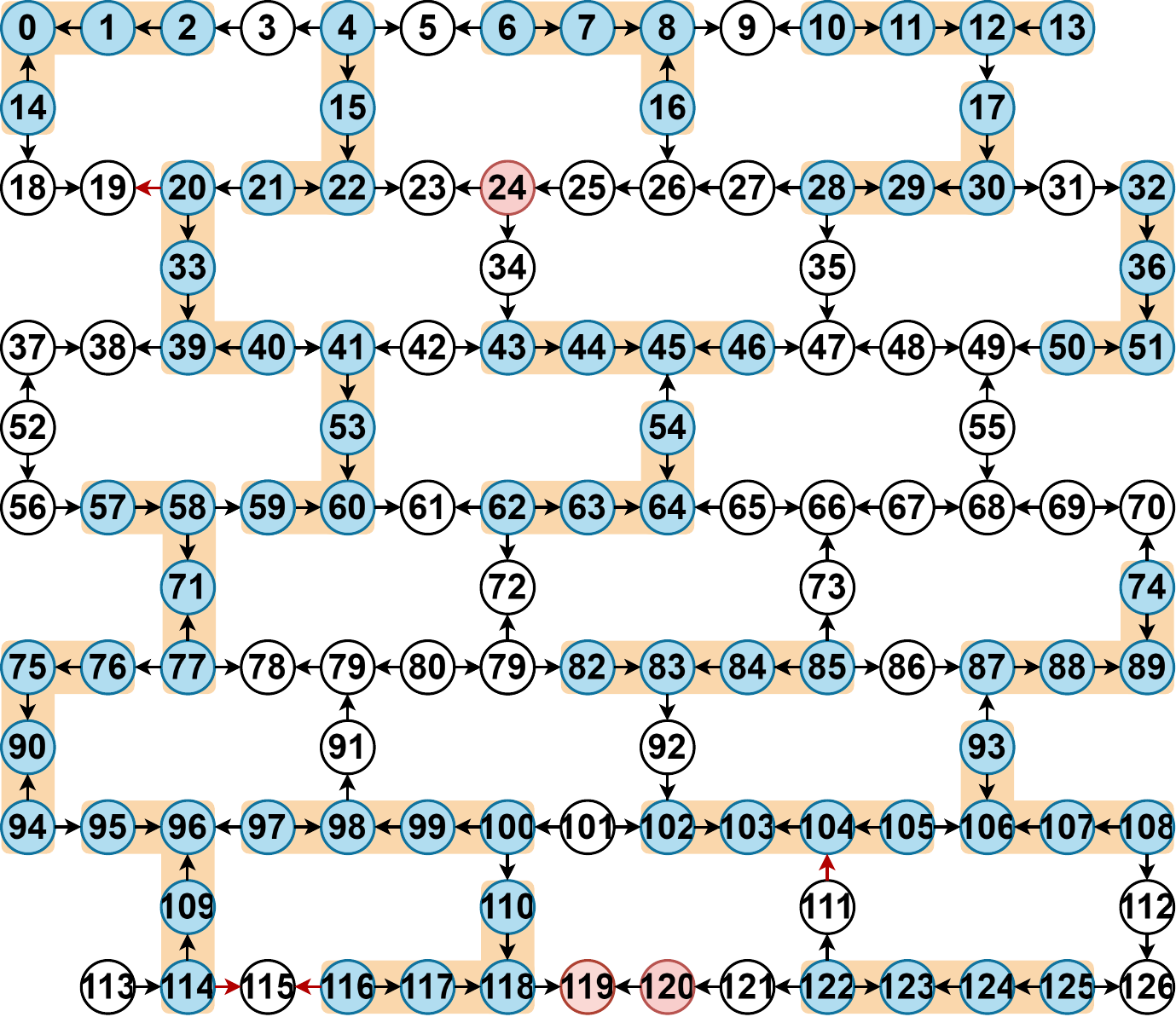}
        \label{fig:brisbane_partition} }
    \\
    \subfloat[Parallelized feature map.]{
        \centering
        \includegraphics[width=0.95\linewidth]{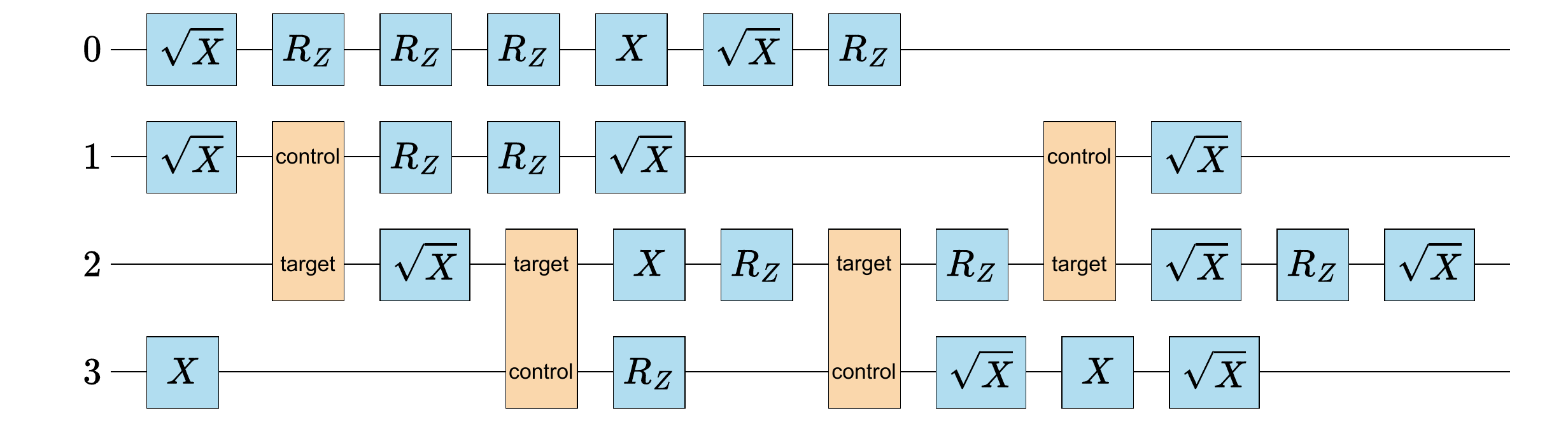}
        \label{fig:fmap} }
    \caption{Partitioning scheme for the ibm\_nazca (a) and ibm\_brisbane (b) $127$-qubits QPUs. Arrows indicate the direction of operation of the native ECR gate. Sub-units are highlighted in orange and the light-blue coloring identifies the qubits used. Red colouring indicates that either qubit readout or ECR gate fidelities were too low to be used. (c) The transpiled, random-generated feature map used in the parallelization experiment, with four ECR gates and a depth of $12$.}
    \label{fig:qpu_partitions}
\end{figure}

\noindent The QPU partitioning for ibm\_nazca and ibm\_brisbane is depicted in Fig.~\ref{fig:qpu_partitions}, in which we carefully selected $21$ sites for ibm\_nazca and $20$ for ibm\_brisbane, sharing the same connectivities, i.e.\ a ``directional chain'' $(\rightarrow \rightarrow \leftarrow)$, where each arrow starts from a control qubit and points to a target qubit.
In choosing the partition schemes, we also avoided qubits with higher readout error rates and gate infidelity, which would negatively affect the kernel output.

\noindent Backend ibm\_torino required a separate analysis, due to the different set of basis gates (I, RZ, SX, X, CZ). The controlled-Z (CZ) interaction, contrarily to ECR for the previous QPUs in our study, can be applied in both directions for any pair of connected qubits. This simplified the partitioning into non-directional spin chain sub-units.

\begin{figure}
    \centering
    \subfloat[Output spread across sites for ibm\_nazca and ibm\_brisbane.]{
        \centering
        \includegraphics[width=0.47\linewidth]{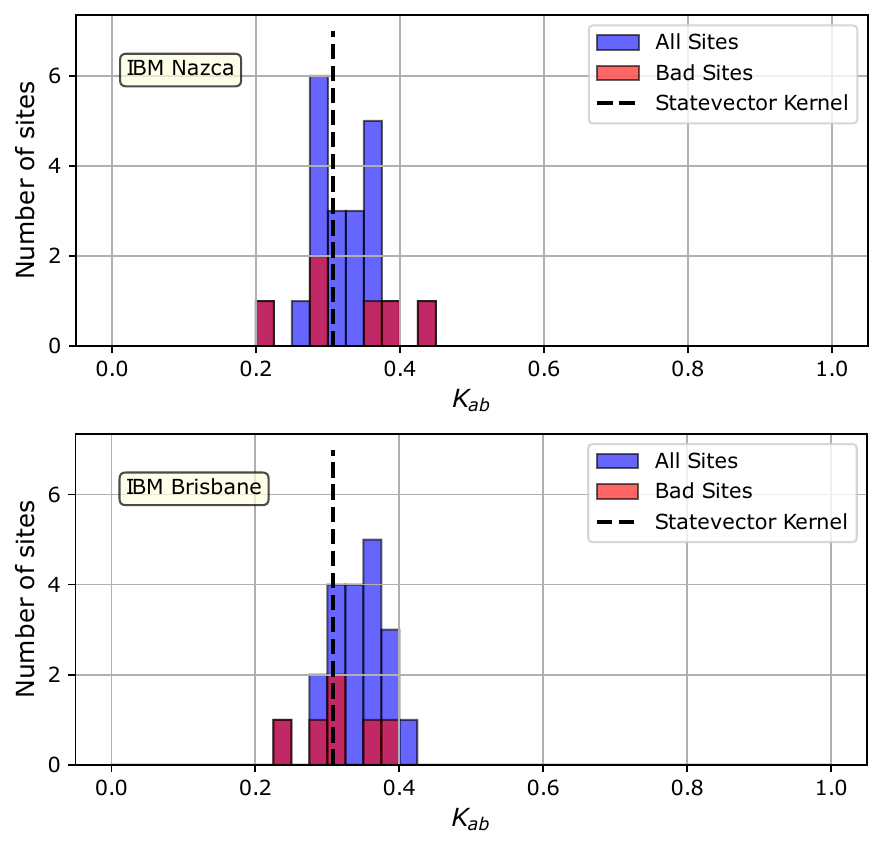}
        \label{fig:output_a}}
    \hfill
    \subfloat[Frobenius norm-based evaluation of ibm\_nazca and ibm\_brisbane sub-units.]{
        \centering
        \includegraphics[width=0.47\linewidth]{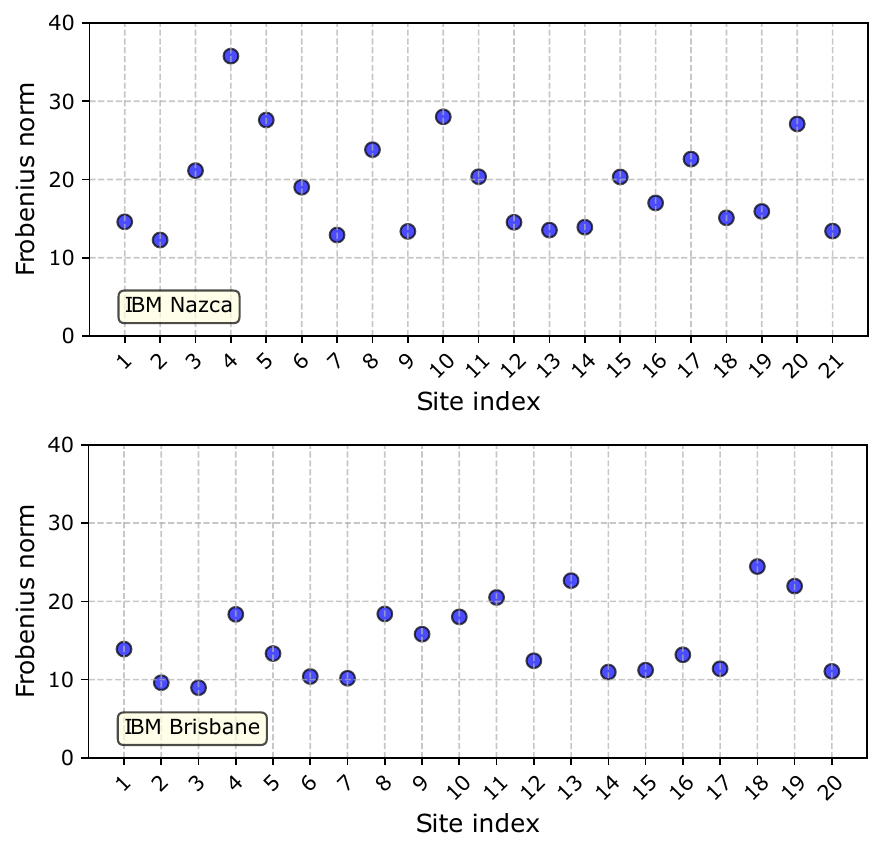}
        \label{fig:output_b}}
    \caption{(a) Output spread across sites for ibm\_nazca and ibm\_brisbane for an example kernel entry. The worst $6$ sites (in terms of Frobenius norm) distributions are coloured in red. The black dashed lines mark the exact value computed with a statevector simulation. (b) Frobenius norm of the difference between the kernel matrix computed in each site and the statevector result.}
    \label{fig:qpu_comparison}
\end{figure}

Running the same quantum kernel evaluation across all QPU sub-units (or sites) for multiple devices allowed us to simultaneously evaluate the impact of noise in terms of output spread, i.e.\ the average standard deviation between the partitions output and the true kernel value. 
We also ranked the sub-units based on their performance, by evaluating the Frobenius norm \cite{frobenius} of the difference between the kernel matrix computed on each sub-unit and the one computed with a statevector simulation. In Fig.~\ref{fig:qpu_comparison} we show an example of kernel entry estimation from all the QPU sub-units and the associated Frobenius norm value for ibm\_nazca and ibm\_brisbane. 


\begin{figure}
    \centering
    \subfloat[Site spread distributions for ibm\_nazca and ibm\_brisbane.]{
        \centering
        \includegraphics[width=1\linewidth]{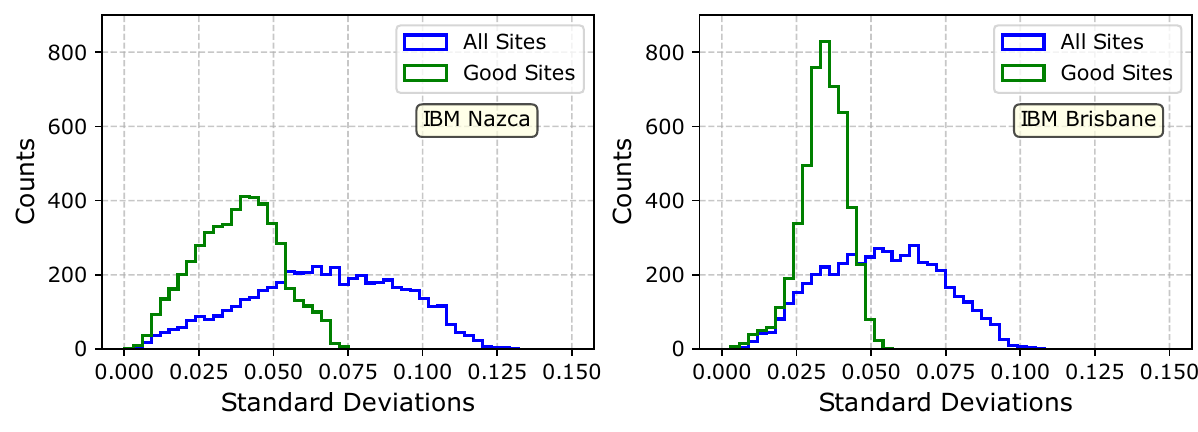}
        \label{fig:spread_a}}
    \hfill
    \subfloat[Average spread distribution as a function of the number of sites excluded for ibm\_nazca and ibm\_brisbane.]{
        \centering
        \includegraphics[width=\linewidth]{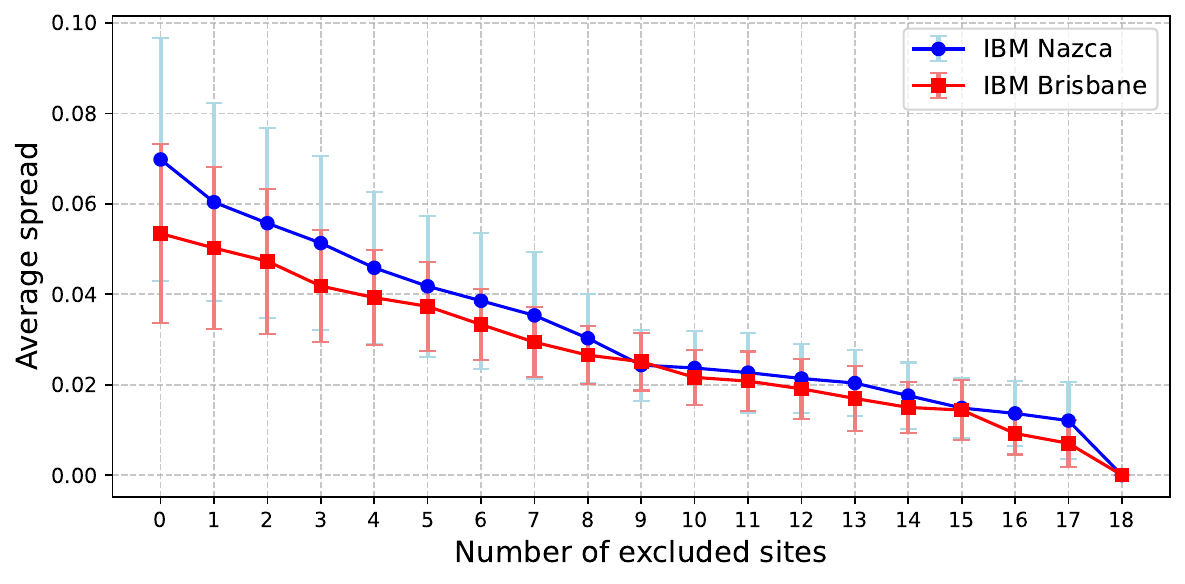}
        \label{fig:spread_b} }
    \caption{(a) Site spread distributions, where each count in a histogram corresponds to the standard deviation of the kernel matrix entries evaluated by each site taken into consideration. The blue histograms collect the results for all partitions, while for the green ones the $6$ worst sites in terms of Frobenius norm were rejected. (b) Average spread, corresponding to the mean value of the distributions in (a), for a different number of excluded sites. Sites were excluded in ascending order of the Frobenius score.}
    \label{fig:nazca_brisbane_comparison_spread}
\end{figure}

Fig.~\ref{fig:spread_a} depicts the distribution of site output spread (standard deviations) for each off-diagonal kernel entry. We then examined how the distributions are affected when progressively discarding the worst sites in terms of the Frobenius norm. In Fig.~\ref{fig:spread_b}  we show how averaging all kernel entries, the backend site spreads decrease proportionally to the number of excluded sites. By construction, when excluding all sites except one, the site spread is zero.

To carry out the extra-platform comparison, we opted to exclude from the analysis one of ibm\_brisbane sites that presented several outliers in the kernel matrix entries, unaligned with the other site outputs, despite exhibiting a low Frobenius score. Moreover, to keep the same number of sites to compare between the 2 backends as a function of the discarded sites number, we also excluded from the analysis two average-performing sites for ibm\_nazca (ranked 11th and 12th out of $21$), so that we could keep a total of $19$ sites both for both QPUs.

From this analysis we observe that the Brisbane output spread is lower, and the spread distribution widths (i.e.\ the error bars in Fig.~\ref{fig:spread_b}) are significantly lower for all the configurations. This resuls is aligned with our choice since ibm\_brisbane exhibits higher single and two-qubit gate fidelities overall. The backend was characterized by a slightly lower Error Per Layered Gate (EPLG), considering the maximum-length qubit chain \cite{eplg}: $3.0\%$ against the $3.3\%$ of ibm\_nazca.

\begin{figure}
    \centering
    \subfloat[Frobenius norm-based evaluation of ibm\_torino partitions.]{
    \centering
    \includegraphics[width=0.8\linewidth]{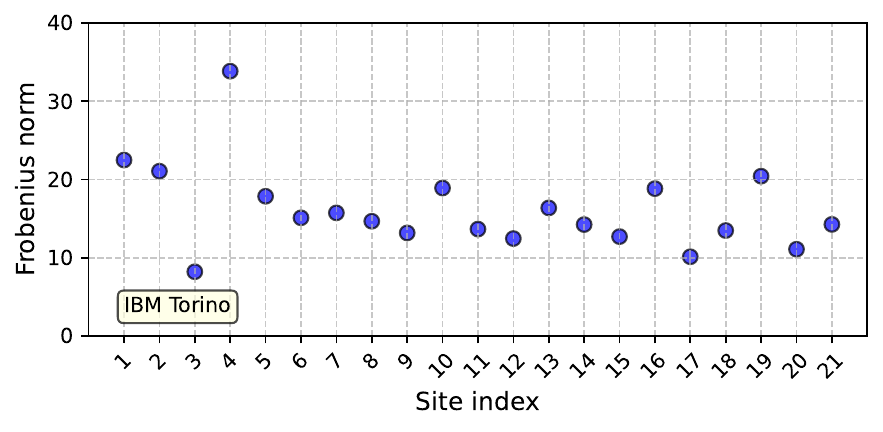}
    \label{fig:torino_a} }

    \subfloat[Average spread distribution as a function of the number of sites excluded for ibm\_torino.]{
        \centering
        \includegraphics[width=0.8\linewidth]{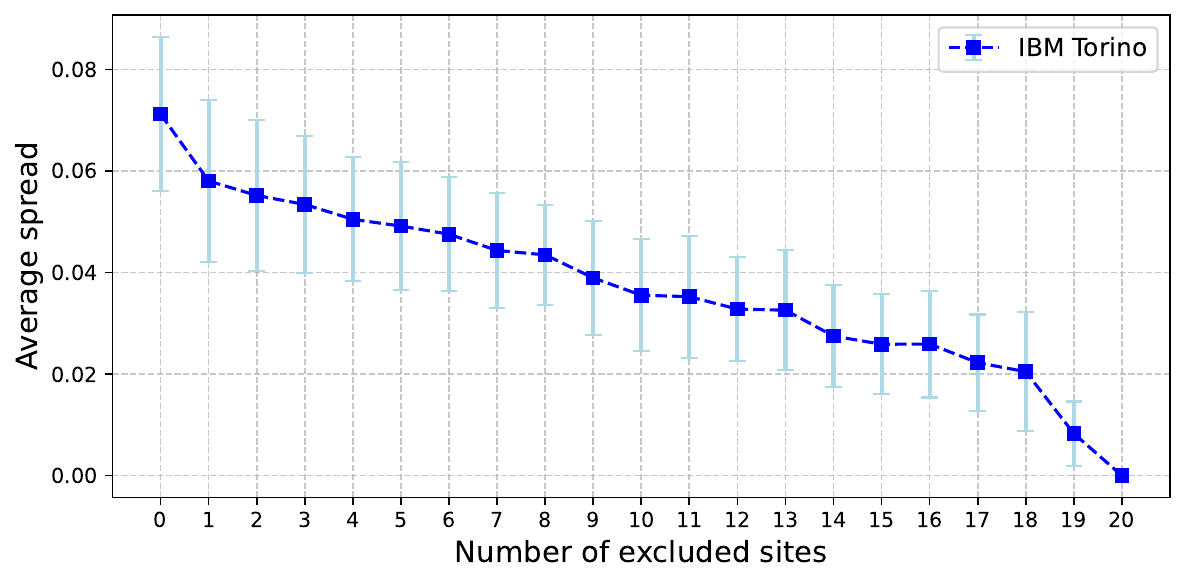}
        \label{fig:torino_b} }
    \caption{(a) Frobenius norm of the difference between the kernel matrix computed in each site and the statevector result. (b) Average site spread as a function of excluded sites number. Sites were excluded in ascending order of the Frobenius score shown in (a).}
    \label{fig:torino}
\end{figure}

We carried out an intra-platform site-spread characterization also for ibm\_torino. For a comparable analysis in terms of feature encoding circuits depth and entangling gates utilized for ibm\_nazca and ibm\_brisbane, we modified the kernel according to the basis gates set of ibm\_torino. Fig.~\ref{fig:torino} shows the output spread analysis for the $21$ designated partitions in the QPU. Despite this platform having the highest two-qubit gate fidelity and lowest EPLG for maximum-length qubit chain ($0.8$) among the QPUs analyzed in this study, the Frobenius distances are comparable with the ones of ibm\_brisbane. 
In addition, the site spread as a function of the worst excluded site number is slightly higher than both ibm\_nazca and ibm\_brisbane. 
This result underlines the impact of noise in different quantum kernel estimations. Our data-driven approach allowed us to estimate the noise impact order of magnitude, which was consistent across the IBM  devices under testing.
Despite the attempt to compute quantum circuits with similar structures, the possibility of precisely quantifying site spread for generic quantum feature maps remains challenging. 

We conclude our study by integrating the site spread information analyzed in this section into statevector genetic optimizations. By simulating the effect of noise directly on the kernel matrix at levels comparable to those observed in NISQ devices, we can assess whether genetic optimization can still be carried out under noisy conditions.

\begin{figure}
    \centering
    \includegraphics[width=\linewidth]{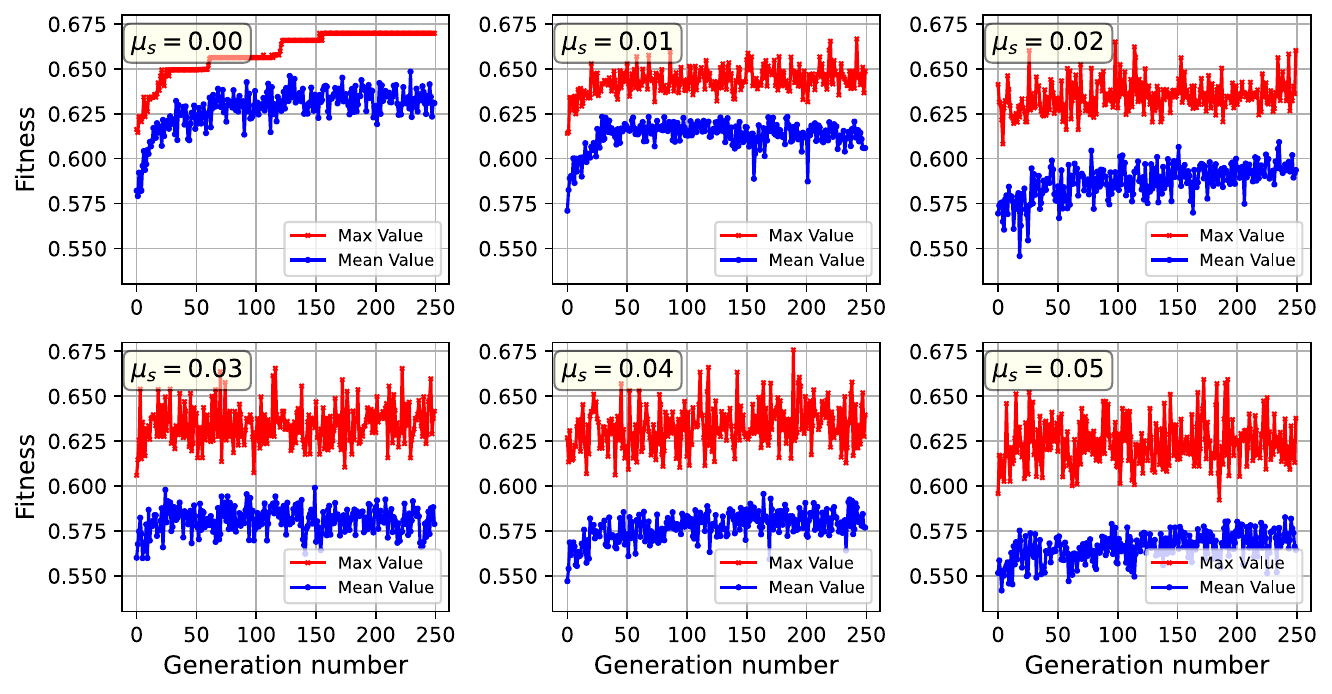}
    \caption{$4$-qubits fitness function through generations via steady-state selection, applying a Gaussian noise with standard deviation $\mu_s$, corresponding to the average site spread. Allowed gates: I, RZ, SX, X, ECR.}
    \label{fig:noisy_simulation}
\end{figure}

\noindent We applied a Gaussian distributed noise with zero mean and different standard deviations to each of the kernel entries. For these genetic optimizations, we considered $4$-qubits quantum kernels with the basis gates set of ibm\_nazca and ibm\_brisbane, and we constrained the two-qubit gate to only apply to qubits that are directly connected in the $\rightarrow \rightarrow \leftarrow$ scheme of sub-units considered in the experiment. The dataset in use was the same as the one analysed in Sec.~\ref{sec:statevec}, and elitism was not applied ($K=0$). The results are shown in Fig.~\ref{fig:noisy_simulation}, exhibiting a positive trend in the average fitness function throughout generations for average site spreads below $\mu_s=0.02$, which is a level within reach of the NISQ devices tested in this study when retaining the best $20\%$ of partitions. The impact of noise on the fittest quantum kernel curve seems more relevant, as a positive trend can be observed only for $\mu_s\leq0.01$. In the $\mu_s=0.01$ case, the best fitness function per generation reaches a peak of $f=0.667$, translating into a cross-validated accuracy of $a=65.8\%$ (following from eq.~\ref{eq:mono}, $\eta = 0.025$ and $\sigma=0.35$). In the ideal case, the fitness function plateaus at $f=0.670$ and $a=66.2\%$.

We expect the fitness function variability to derive both from noise and feature map variability throughout generations. This suggests that noise will impact differently in proportion to the overall variability for different genetic settings such as mutation and crossover rates. The effect of noise is also related to the circuit depth and the overall noise dependence of the other metrics composing the fitness function. We also expect the dataset size to play a role in the noisy optimization, as the intrinsic cross-validated accuracy fluctuations would be more relevant when considering small-sized datasets.

\section{Conclusions}
\label{sec:conclusions}
In this study, we introduced a novel feature map optimization strategy utilizing both mono-objective and multi-objective genetic algorithms to enhance the performance of QSVMs for binary classification tasks, with a specific focus on applications in neutrino physics. Our approach significantly improved the classification accuracy for single and double beta decay topologies ($\beta$ and $\beta\beta$), as the genetic algorithm consistently converged to feature maps yielding higher accuracies than the randomly initialized ones. Notably, our optimized QSVMs systematically outperformed classical Gaussian kernel SVMs, suggesting that the problem can be addressed more efficiently with quantum algorithms or quantum-inspired methods, under the same dataset dimensionality. 
Furthermore, despite the performance of deep learning methods for this task \cite{Moretti2024} remains unmatched for orders of magnitude larger datasets, the rarity of double beta decay events motivates the development of classification strategies capable of achieving high performance with limited training data. 

From a pure algorithmic perspective, we explored a novel and potential approach for parallelizing genetic searches directly on quantum hardware, by executing identical quantum kernel circuits simultaneously on dedicated $4$-qubits spin chain-like QPU partitions. These experiments allowed us to evaluate the noise-driven output spread on the kernel function values across such partitions. Incorporating this data into statevector simulations allowed us to estimate an output spread threshold above which the fitness function stops improving throughout generations. Our findings suggest that, after careful selection of the best-performing partition, parallelized genetic optimization on quantum hardware is within reach of the tested devices.

\noindent While this parallelization effort has shown potential, several near-term challenges remain. These include upscaling the qubit number in QPU partitions and exploring more complex connectivity schemes to optimize the distribution to multiple chips (extra-platform execution). For instance, significant benefits could be gained by including an automated transpilation process of the overall circuit, consisting of all sub-units, into an equivalent one with optimized depth, also taking into account readout error and gate fidelities of individual qubits. In this work we did not include and discuss quantum error mitigation strategies into the algorithm, which is left for future work.

Our work highlights the need for further exploration of quantum parallelization strategies and deeper integration with hardware-aware compilation techniques in the utility scale regime and provides useful insights for future studies aiming to enhance the practical utility of QML in real-world applications.

\vspace{1cm}
\textbf{Acknowledgements}
This work is supported by PNRR MUR projects PE0000023-NQSTI and CN00000013-ICSC and by QUART\&T, a project funded by the Italian Institute of Nuclear Physics (INFN) within the Technological and Interdisciplinary Research Commission (CSN5) and Theoretical Physics Commission (CSN4). MG is supported by CERN through CERN Quantum Technology Initiative. We gratefully acknowledge Matteo Biassoni for generating the dataset used throughout the study. We also thank Daniele Guffanti and Francesco Terranova for many insightful discussions on LArTPCs and the DUNE low-energy physics. We acknowledge the use of IBM Quantum services for this work. The views expressed are those of the authors, and do not reflect the official policy or position of IBM or the IBM Quantum team.
\vspace{1cm}

\textbf{Data availability statement}
The dataset generated and analysed during this study is available upon reasonable request from the author. The manuscript has associated data in a repository: \href{https://github.com/rmoretti9/MaQIP.git}{https://github.com/rmoretti9/MaQIP.git}

\vspace{1cm}
\textbf{References}

\bibliographystyle{unsrt}

\bibliography{bibl}
\end{document}